\documentstyle[pra,aps,preprint]{revtex}
\begin{document}
\draft
\title{Kondo Effect in High-T$_{\rm c}$ Cuprates }
\author{Naoto Nagaosa}
\address{Department of Applied Physics, University of Tokyo,
Bunkyo-ku, Tokyo 113, Japan}

\author{Patrick A. Lee}
\address{Department of Physics, Massachusetts Institute of Technology,
Cambridge, MA02139, U.S.A.}
\date{\today}
\maketitle
\begin{abstract}
We study the Kondo effect due to the nonmagnetic
impurity, e.g., Zn, in high-T$_{\rm c}$ cuprates
based on the spin-change separated state.
In the optimal or overdoped case with the Kondo screening,
the residual resistivity is dominated by the spinons
while the T-dependent part determined by the holons.
This gives $\rho(T) = { {4 \hbar} \over {e^2}}
{ { n_{\rm imp.}} \over {1-x}}  + {{\alpha T} \over x}$
($x$: hole concentration,$n_{\rm imp.}$: impurity concentration,
$\alpha$: constant ), which is in agreement with experiments.
In the underdoped region with the pseudo spin gap,
an SU(2) formulation predicts that the holon phase  shift is
related to the formation of the local spin moment, and hence the
residual resistivity is given by
$\rho_{\rm res.} = { {4 \hbar} \over { e^2}}
{ { n_{\rm imp.}} \over {x}} $, which is also consistent with the
experiments.
The magnetic impurity case, e.g., Ni, is also discussed.
\end{abstract}

\narrowtext
Kondo effect is a phenomenon shown by a magnetic impurity put
into a nonmagnetic metal \cite{kondo}.
As the temperature is decreased, the magnetic moment is screened by the
conduction electrons and finally the Kondo singlet, i.e.,
the singlet of the localized spin and the conduction
electrons, is formed.  In the usual Kondo effect, the conduction
electrons are assumed to be non-interacting or to form a Fermi liquid,
which is magnetically inert due to Fermi degeneracy.
When the conduction electrons are strongly correlated
and magnetically active, it is expected that Kondo effect
is also modified.
High-T$_{\rm c}$ cuprates offer a unique opportunity to study such
an effect.  In the undoped high-Tc cuprates, the valency
of Cu atom is Cu$^{2+}$ ($d^9$)
and the system is a Mott insulator.   By the hole doping, the system becomes
metallic and shows superconductivity with high Tc.
We believe that the  Kondo effect in this system is actually observed for the
{\it nonmagnetic} impurity, e.g., Zn,
replacing Cu atom in the conducting plane.
The valency of Zn is Zn$^{2+}$ ($d^{10}$)
and compared with the Cu$^{2+}$
case one electron is trapped by one additional positive charge of the neucleus,
which forms a singlet on the Zn site.
In the underdoped cuprates with spin gap, it is found experimentally that
a local moment of $S=1/2$
appears on neighboring Cu sites
\cite{xiao,alloul,zheng,ishida,mahajan,mendels,fin,ong,uchida1,uchida2}.
We believe this localized spin is not screened by
the conduction electron spins
because of the reduced density of states for spins at the
Fermi energy $E_F$ in the presence of the spin gap \cite{taikai,kal}.
Once the spin gap collapse with the
increased hole concentration, the density of states for spins at $E_F$
recovers and also the Kondo screening, i.e.,
the singlet formation between localized spin and conduction
spins, occurs.
Associated with the formation of local moments, it is found that
the residual resistivity $\rho_{\rm res.}$ is very large in high-Tc cuprates.
For example  $\rho_{\rm res.}$ at 1$\%$ Zn doping
in La$_{2-x}$Sr$_x$CuO$_4$ ($x=0.15$) amounts to $\sim 100 \mu \Omega cm/\%$.
This value should be compared with
$ \rho_{\rm res.} = 0.32 \mu \Omega cm/\% $ for Zn doping in the
Cu metal \cite{ziman}. The latter is understood in terms of the Born
approximation using the screened Coulomb potential, and hence
$\rho_{\rm res.}$ is proportional to $Z^2$ ($Z$: the difference of the valence
between the host and impurity atoms) as observed
for Zn, Ga, Ge, As in Cu metal, and this explains the small
$\rho_{\rm res.}$ for Zn with $Z=1$ \cite{ziman}.
Because the d-orbitals of the impurity atoms are completely occupied
in these cases, the resonant scattering is absent and the
phase shifts are distributed to various partial wave components $\ell$.
 On the  other hand, d-obitals of Cr, Mn,  Fe, Co, Ni atoms in Cu metal are
partially filled, and cause the resonant scattering.
Then $\rho_{\rm res.}$ is dominated by the d-wave component
$\ell=2$, and the residual resistivity can be
analysed in terms of the Friedel sum rule \cite{friedel} and the Kondo effect.
Friedel sum rule is the expression of the charge neutrality,
and is given explicitly by
\begin{equation}
Z = { 1 \over \pi} \sum_{\ell,\sigma} ( 2 \ell + 1) \delta_{\ell, \sigma}
\end{equation}
where
$\delta_{\ell, \sigma}$ is the phase shift for the partial wave component
$\ell$ with spin $\sigma$.
Let $S$ be the spin of the impurity, and
the phase shifts are given by
\begin{eqnarray}
Z &=& { {2 \ell +1} \over \pi} ( \delta_{\uparrow} + \delta_{\downarrow} )
\nonumber \\
2S &=& { { 2 \ell +1} \over \pi} ( \delta_{\uparrow} - \delta_{\downarrow} )
\end{eqnarray}
with $\ell = 2$.
The spin $S$ becomes zero below the Kondo temperature
$T_K$ due to Kondo screening,
and these two equations determine the phase shifts
and hence the residual resistivity below and above $T_K$ \cite{shiba}.
This prediction is consistent with the experiments.
For examples Fe in Cu ( $Z$ = 3 ) shows
$\rho_{\rm res.} = 18.5 \mu \Omega cm/\% $ below $T_K$
corresponding to $S=0$ and $Z=3.3$,
whcih is more than 4 times larger than the
Si case  ($ \rho_{\rm res.} = 3.95 \mu \Omega cm/\% $ with $Z=3$).
Then it is generally true that the  larger residual resistivity is
expected when only one $\ell$ component contributes.

Now let us consider the case of high-T$_{\rm c}$ cuprates. It has been
convincingly  discussed that the degeneracy of the d-orbitals is lifted
by the crystal field and only d$_{x^2-y^2}$ orbital is relevant to the
conduction. Furthermore, the single band $t$-$J$ model is
the low energy effective model \cite{rice}. Then it is expected that the s-wave
($\ell=0$) scattering dominates the residual resistivity.  Assuming s-wave
scattering, the  residual resistivity
in the limit of small $n_{{\rm imp.}}$ is given in two dimensions as
\begin{equation}
\rho_{\rm res.} = { {2\hbar} \over {e^2}}
{ { n_{\rm imp.}} \over n} (\sin^2 \delta_{0 \uparrow} +
\sin^2 \delta_{0 \downarrow} )
\end{equation}
where $n_{\rm imp.}$ is the impurity concentration,
$n$ the carrier concentration, and $\delta_{0 \sigma}$
is the phase shift \cite{ong,ziman}.
Theoretically it is not  a trivial problem whether the carriers are the
electrons with the concentration $n=1-x$ or the doped holes with $n=x$.
The experimentally observed values of
$\rho_{\rm res.}$ in the underdoped region are
quantitatively fitted by eq.(3) by putting
the carrier concentration $n=x$, and the phase shift $\delta_{0 \sigma}=\pi/2$
( unitarity limit) \cite{ong,uchida1,uchida2}.
Note that this is the largest value theoretically expected form eq.(3).
As the doping proceeds to the optimal and overdoped regions,
the residul resistivity decreases and fits the formula eq.(3) with
$n$ being increasing to $1-x$ with $\delta_{0 \sigma}=\pi/2$ unchanged
\cite{uchida2}.
This crossover seems to correspond to the disappearance of the local moment.
It is noted that $\rho_{\rm res.}$ in the optimal and overdoped
regions is consistent with the Fermi liquid picture, where
$S=0$ and $Z=1$ in eqs. (2) and (3).
However considering the fact that the conductivity without the
impurities is proportional to $x$ and hence is dominated by the hole
carriers in the optimal doping region, the residual resistivity
coresponding to $n=1-x$ is a mystery.
Even more unconventional is the underdoped case, where the
phase shift is not for the electrons.  Can one consider the phase shift
for the holes ?  Then what determines that phase shift ?
These are the questions to which we give solutions below.

The transport properties in high T$_{\rm c}$ cuprates have been analyzed
in terms of the gauge model based on the mean field theory
of RVB states \cite{bas,nagaosa,ioffe}.
Let us first consider the U(1) theory which is applicable to optimally
doped and overdoped regions.
In this formalism the electron ($C^\dagger_{i \sigma}$)
is described as the composite particle of spinon
($f^\dagger_{i \sigma}$) and holon ($b_{i}$), i.e.,
\begin{equation}
C^\dagger_{i \sigma} = f^\dagger_{i \sigma} b_i
\end{equation}
with the constraint
\begin{equation}
\sum_{\sigma} f^\dagger_{i \sigma} f_{i \sigma}
+ b^\dagger_i b_i = 1.
\end{equation}
This constraint is taken care of by the gauge field.
When the external potential is introduced, the time component
is induced so that eq.(5) is satisfied.
We recall that the resistivity is given by the Ioffe-Larkin composition rule
$\rho=\rho^{{\rm spinon}}+\rho^{{\rm holon}}$ \cite{ioffe}.
For the clean case it is
dominated by $\rho^{{\rm holon}}$ which is inversely proportional to $x$.
The residual resistivity $\rho_{\rm res.}$
is given as
\begin{equation}
\rho_{\rm res.} =
\rho_{\rm res.}^{\rm spinon}+
\rho_{\rm res.}^{\rm holon}  =
{ {4\hbar} \over {e^2}} n_{\rm imp.}
\biggl[
{ { \sin^2 \delta^{\rm spinon}} \over {1-x} } +
{ { \sin^2 \delta^{\rm holon}} \over {x} } \biggr].
\end{equation}
Assuming that the local moment is screened so that the impurity is
nonmagnetic, we conclude that
$\delta^{{\rm spinon}} =\delta_0 = \pi/2$
just by applying the discussion for the Kondo effect for electrons to spinons.
Note that the factor of 1/2 in $\delta_0$ comes from the spin degeneracy
for the spinons and electrons.
For holons, on the other hand, this factor does not occur because
there is only one species of holons.
Then the phase shift of the holons is an integer multiple of $\pi$.
This gives
\begin{equation}
\rho_{\rm res.} = { {4 \hbar} \over { e^2}}
{ { n_{\rm imp.}} \over {1-x}}
\end{equation}
in agreement with the experiments.
As noted before, the appearance of $1-x$ in $\rho_{\rm res.}$
when $x$ appears in the $T$-dependent part of $\rho$ is highly
nontrivial and may be regarded as an
important test of the Ioffe-Larkin rule.

There is one catch in the above argument.
In the case of fermions which is Fermi-degenerated, the phase
shifts appearing in eqs.(3) and (6) are those at the Fermi energy
$E_F$.
This is because both the fermion number and the resistivity
is determined by the integral including
$- \partial f(E)/\partial E \cong \delta( E-E_F)$
($f(E)$: Fermi distribution function).
However in the case of bosons,
$- \partial n(E)/\partial E $
($n(E)$: Bose distribution function)
gives a rather broadened weight in the integral and
$\delta^{\rm holon}$
can not be replaced by the value at a respresentative energy.
This is serious especially at low energy where
$\delta^{\rm holon} \propto k \propto E^{1/2}$.
Actually the holons are hard-core bosons which interact strongly
with the gauge field. It is expected that the gauge
fluctuation will reduce the distribution for
$k \cong 0$ to make $n(E)$
flatter as a function of $E$.
Another clue to the functional form of $n(E)$
is that the hard-core bosons are in a sense similar to the
fermions as suggested from the 1D models.
From these we assume that
$- \partial n(E)/\partial E $
is sharply peaked enough to replace $\delta^{\rm holon}$
by a representative value.

Next we discuss the underdoped regime. From our discussion up to now, it
is very difficult to explain the experimental observation. Since the
resistivity is proportional to $x$, we need the holon scattering to be
unitary. However, since there is no spin label on the holon,
the natural values
for $\delta^{{\rm holon}}$ is 0 or $\pi$, and in either case
$\rho_{\rm res.}^{\rm holon}$ is zero.
We also observe that unlike the overdoped case, where the Kondo scattering is
smoothly connected to the strongly overdoped Fermi liquid limit,
in the underdoped case we cannot recover the experimental value even if we
extrapolate to the zero doping limit, i.e., Neel state.
In this case the unit cell is doubled, and the doped holes form two
inequivalent Fermi pockets near $(\pi/2,\pm \pi/2)$.
Let us assume that a local moment is formed, an assumption which is by no means
obvious. By extending eqs.(1) and (2) to include 2 pockets and setting
$Z=1$, $S=1/2$, we find $\delta_{\uparrow} =\pi/2$ and
$\delta_{\downarrow} = 0$.  From eq.(3) we find
$\rho_{\rm res.} = { {2 \hbar} \over { e^2}}
{ { n_{\rm imp.}} \over {x}} $, which is  still small compared with the
experiment by a factor of 2. On the other hand, if no local moment is formed,
we find $\delta_{\uparrow} = \delta_{\downarrow} = \pi/4$ and
$\rho_{\rm res.}$ is even smaller. Thus it is apparent that the
experimental observation is highly nontrivial to explain.

 Recently, it was pointed out that the traditional formulation of the
 $t$-$J$ model ( which we shall call the U(1) formulation ), is
 inadequate for small doping, because it does not include low lying
 excitations connected to the SU(2) symmetry which is known to exist
 exactly at half filling \cite{affleck}. A new formulation was introduced
 which includes that fluctuations, and it is believed to be a better
 starting point for the underdoped region \cite{su2}. A feature of the SU(2)
 formulation is that two bosons $(b_1,b_2)$ which form an SU(2) doublet
 is introduced. Instead of eq.(5), the constraint is given by
\begin{equation}
\sum_{\sigma} f^\dagger_{i \sigma} f_{i \sigma}
+ b^\dagger_{1i} b_{1i} - b^\dagger_{2i} b_{2i} = 1
\end{equation}
 and the number of vacancy is given by
$b^\dagger_{1i} b_{1i} + b^\dagger_{2i} b_{2i}$.
%Note that the SU(2) spinon $f^\dagger_{i \sigma}$,$f_{i \sigma}$ is
%different from the U(1) fermion in eqs.(4) and (5).
%However the expression for the spin
%${\vec S}_i$ remains the same, i.e.,
%${\vec S}_i = { 1 \over 2} f^\dagger_{i \alpha} {\vec \sigma}_{\alpha \beta}
%f_{i \beta}$.
We recover the U(1) formulation if the boson isospin doublet $(b_{1i},b_{2i})$
is polarized in the $z$ direction, giving $(b_i,0)$.  In contrast, in the SU(2)
mean field theory \cite{su2}, the underdoped normal state is represented by the
staggered flux phase, where the constraint is satisfied by $\langle
b_{1}^\dagger b_{1} \rangle = \langle b_{2}^\dagger b_2 \rangle $, i.e. the
isospin is strongly fluctuating.  We shall argue that the  experiment may be
explained using this new formulation. The first step is to  remove the Zn site
from consideration by treating it as a vacancy in the  Cu(d$^9$) lattice.
This is
reasonable because  Zn(d$^{10}$) is charge neutral relative to Cu(d$^9$) and is
a spin singlet. We can model the vacancy by a strong repulsive local potential
for spinon  and holon. Next we argue that the potential will form a spinon bound
state,  which is responsible for the local moment. This has been shown to be the
case in the spin gap phase if the spinon spectrum has point nodes, and a linear
density of states \cite{taikai,kal}. Now we treat all the Cu sites using the
SU(2) formulation. Within a large sphere, the formation of the local spin means
that the  number of spinons has increased by 1, i.e.,
$\Delta( \sum_{i: {\rm in \  the \  sphere}}
f^{\dagger}_{i \sigma} f_{i \sigma}) = 1$. Using the constraint eq.(8)
we see that
\begin{equation}
\Delta \sum_i(
 b^\dagger_{1i} b_{1i} - b^\dagger_{2i} b_{2i}) = -1.
\end{equation}
At the same time, charge neutrality requires that
\begin{equation}
\Delta \sum_i(
 b^\dagger_{1i} b_{1i} + b^\dagger_{2i} b_{2i}) = 0.
\end{equation}
The only way to satisfy eq.(9) and (10) is for
\begin{equation}
\Delta \sum_i(b^\dagger_{1i} b_{1i}) =
-\Delta \sum_i(b^\dagger_{2i} b_{2i}) = -{ 1 \over 2}.
\end{equation}
 In terms of phase shift, we have
 $\delta_{b_2} = - \delta_{b_1} = \pi/2$, leading to a residual resistivity
 of
\begin{equation}
\rho_{\rm res.} = { {4 \hbar} \over { e^2}}
{ { n_{\rm imp.}} \over {x}}
\end{equation}
in agreement with experiment.
We note that our argument so far 
applies to any divalent nonmagnetic impurity, such as Zn and Mg.
 Let us now consider a magnetic impurity such as Ni.
In this case the Ni is in a d$^8$ configuration with $S=1$. In the
underdoped case, the argument proceeds as before, except that the additional
$S=1/2$ on the bound state will have strong antiferromagnetic
exchange with the $S=1$, leading to an $S=1/2$ local moment. The boson
counting is the same as before, and we predict eq.(12) to hold. This
is in fact the experimental situation \cite{ni,ong2}.
In the optimal or overdoped case, we
believe the $S=1$ moment will be partially screened, so that
an $S=1/2$ local moment remains. This can be viewed also as a ferromagnetic
Kondo problem, as shown by Khaliullin et al. \cite{kal}. In this case we expect
 $\delta_{\uparrow} =  \delta_{\downarrow} = 0$,
 and the spinon scattering should be very weak. Indeed, experimental
 $\rho_{\rm res.}$ is much smaller that the Zn doped case in the optimal or
 overdoped case \cite{ong2}.

 To complete the discussion, we have to argue that it is plausible to assign
 a phase shift to the $b_1$ and $b_2$ bosons, as if they were fermions.
In the SU(2) formulation, the effective Lagrangian
describing the holons $h_i = [ b_{1i}, b_{2 i}]$
in the underdoped spin gap region is given by \cite{su2}
\begin{eqnarray}
L &=& \int d r h^{\dagger} ( r, \tau)
[ \partial_\tau + i a_0^3 \tau_3 + i A_0
+ { 1 \over {2m}}( - i \nabla +  {\vec a}^3 \tau_3 + {\vec A})^2 - \mu ]
h(r,\tau)
\nonumber \\
&+& \sum_{q,\omega} a^3_{\mu}(q,\omega)
\Pi_{S \mu \nu}(q,\omega)a^3_{\nu}(-q,-\omega)
\end{eqnarray}
where the spinons have been integrated over to give the
polarization function $\Pi_S$ in the effective acton for the gauge field.
Because the gauge symmetry is broken from SU(2) to U(1) in the
staggered flux state, only $a^3$ gauge field remains massless.
Note also that  the Ioffe-Larkin composition
rule no longer applies because the
external vector potential $A_{\mu}$ is coupled to $h_i$ with the
unit matrix and not with $\tau_3$.
Then the conductivity is totally determined by that of
the holon system.
Here we have the two problems, i.e., the strong gauge field fluctuation
and the hard core condition for the holons.
We believe that these two are resolved simultaneously by introducing the
statistical transmutation of $b_1$,$b_2$ to fermions.
This is accomplished by introducing the Chern-Simons gauge field
$a'$ coupled to $b_1$,$b_2$ with the $+$ and $-$ gauge charges,
respectively \cite{scz}.
\begin{eqnarray}
L &=& \int d r h^{\dagger} ( r, \tau)
[ \partial_\tau + i (a_0^3 + a'_0)\tau_3 + i A_0
+ { 1 \over {2m}}( - i \nabla +  ({\vec a}^3 + {\vec a'} ) \tau_3 +
{\vec A})^2 - \mu ]
h(r,\tau)
\nonumber \\
&+&  a^3 \Pi_{S} a^3
+  (a^3 + a') \Pi_{H} (a^3 + a')
+  a' \Pi_{CS} a'
\end{eqnarray}
In the Coulomb gauge, gauge field has two components
as $a_0$ and $a_1 = a_{\rm transverse}$. In this representation,
$\Pi_{CS}$ is given by
\begin{equation}
(\Pi_{CS})_{01} =
(\Pi_{CS})_{10} = c q =  { q \over {2 \theta}}
\end{equation}
with the diagonal components being zero.
Here $\theta$ is the statistical angle, and the bosons are transformed
into fermions when $\theta = (2m+1)\pi$ ($m$: an integer).
We take this choice because the hard core condition is automatically taken
into account even for the noninteracting fermions.
Because of the opposite charges of $b_1$ and $b_2$, the
system remains gauge neutral and also the time reversal symmetry is
preserved at the mean field level.
Now the holon $h_i$ is coupled to $(a^3 + a')\tau_3$,
and we obtain the effective action for
$a= a^3+a'$ by integrating over $a^3-a'$.
Then the gauge flux fluctuation $D = <(\nabla \times a) \cdot
( \nabla \times a)>$
is given by
$D = { 1 \over { \Pi_{H1} + \Pi_{S1}} } f$
with the factor $f$ being
\begin{equation}
f = {
{ c^2 q^2(\Pi_{H0} + \Pi_{S 0}) - \Pi_{H0} \Pi_{S0} \Pi_{S1} }
\over
{ c^2 q^2(\Pi_{H0} + \Pi_{S 0}) -
\Pi_{H0} \Pi_{H1}\Pi_{S0} \Pi_{S1}/(\Pi_{H0} + \Pi_{S0} ) } }
\end{equation}
$\Pi_{\alpha 0}$ and $\Pi_{\alpha 1}$ are the diagonal
longitudinal and transverse components of the
spinon ($\alpha = S$) and holon ($\alpha = H$) polarization function.
It is easy to see that $f<1$ and this factor represents the
reduction of the gauge field fluctuation.
The physical picture is that large part of the original gauge field
is cancelled by the Chern-Simons gauge field attached to the fermions
by an appropriate choice of the integer $m$.
Then it is expected that the hard-core potential and the strong
gauge field fluctuations are taken into account in terms of the
two-component free fermion theory, and this
may justify the phase shift argument given above.

In summary we have analyzed the Kondo effect in high-T$_{\rm c}$ cuprates
based on the spin-change separated state.
The phase shift $\delta^{\rm holon}$ ($\delta^{\rm spinon}$)
for holons (spinons) changes from  $\pi/2$ to 0
( 0 to $\pi/2$ ) due to the Kondo screening together with the crossover
from SU(2) to U(1) theory, which explains the
change of the  residual resistivity from
$\rho_{\rm res.} = { {4 \hbar} \over { e^2}}
{ { n_{\rm imp.}} \over {x}} $
to
$\rho_{\rm res.} = { {4 \hbar} \over { e^2}}
{ { n_{\rm imp.}} \over {1-x}} $
as the hole concentration increases and the local moment disappears.
Lastly we comment on the bipolaronic model for the
underdoped cuprates.   In this model the charge of carrier is
$2e$ and $n = x/2$ in the underdoped region.
This gives
$\rho_{\rm res.} = { {2 \hbar} \over { e^2}}
{ { n_{\rm imp.}} \over {x}} $, which is half
of that expected above.   Then the experiments support the
existence of the carrier not with change $2e$  but with $e$.

The authors acknowledges S.Uchida, N.P.Ong, Y.Kitaoka, H.Takagi,
for fruitful discussions. N.N.  is supported by 
Priority Areas Grants 
and Grant-in-Aid for COE research
from the Ministry of Eduction, Science and Culture of Japan,
P.A.L acknowledges the support by NSF-MRSFC grant number DMR-94-00334.

\end{document}